\documentclass[prl,aps,showpacs,twocolumn]{revtex4}%
\usepackage{graphicx}
\usepackage{amsmath}
\usepackage{amsfonts}
\usepackage{amssymb}%
\providecommand{\U}[1]{\protect\rule{.1in}{.1in}}
\providecommand{\U}[1]{\protect\rule{.1in}{.1in}}
\providecommand{\U}[1]{\protect\rule{.1in}{.1in}}
\providecommand{\U}[1]{\protect\rule{.1in}{.1in}}
\begin{document}
\title{Self-Assembly of Colloidal Superstructures in Coherently Fluctuating Fields}
\author{Igor M. Kuli\'{c}$^{1}$}
\email{kulic@unistra.fr}
\author{Miodrag L. Kuli\'{c}$^{2}$}
\email{kulic@th.physik.uni-frankfurt.de}
\date{\today}

\begin{abstract}
>From microscopic fluid clusters to macroscopic droplets, the
structure of fluids is governed by the Van der Waals force, a
force that acts between polarizable objects. In this Letter, we
derive a general theory that describes the non-equilibrium
counterpart to the Van der Waals force, which emerges in spatially
coherently fluctuating electromagnetic fields. We describe the
formation of a novel and complex hierarchy of self-organized
morphologies in magnetic and dielectric colloid systems. Most
striking among these morphologies are dipolar foams - colloidal
superstructures that swell against gravity and display a high
sensitivity to the applied field. We discuss the dominance of many
body forces and derive the equation of state for a material formed
by the coherent Van der Waals force. Our theory is applied to
recent experiments in paramagnetic colloidal systems and a new
experiment is suggested to test the theory.

\end{abstract}
\affiliation{$^{1}$CNRS, Institute Charles Sadron, 23 rue du Loess BP 84047, 67034
Strasbourg, France}
\affiliation{$^{2}$ Institute for Theoretical Physics, Goethe-University D-60438 Frankfurt
am Main, Germany }

\maketitle

The central goal of physics is to understand and control the
forces of nature. In recent decades scientists have begun to
invent ways to combine the fundamental physical forces and to
generate new, effective interactions on microscopic and
macroscopic scales \cite{ColdAtoms,Colloidal Suspensions,
Levitron,Block-Copolymers,Labyrinthine}. By utilizing effective
interactions cold atoms can now be trapped and cooled
\cite{ColdAtoms}, colloidal suspensions can be stabilized
\cite{Colloidal Suspensions}, and magnetic levitation has become
possible \cite{Levitron}. In condensed matter physics, the
interplay of different attractive and repulsive forces can give
rise to highly complex structures, ranging from gyroid phases in
block copolymers \cite{Block-Copolymers}, to labyrinthine phases
in ferrofluids \cite{Labyrinthine} and nuclear pasta phases in
neutron stars \cite{Nuclear-Pasta}. Not surprisingly, every
additional interaction increases the complexity of the resulting
materials. But complex materials can also emerge from a
\emph{single}, simple to generate, effective interaction.

In this Letter we explore one of the simplest effective
interactions that is able to generate complex structures. This
interaction is the \textit{spatially coherent fluctuation
interaction} (scFI). The interaction occurs between dipolar
magnetic (or dielectric) particles when a spatially uniform
electromagnetic field varies in time (cf. Fig.1b). The first
realization of this interaction was described by Martin et al. in
a system of superparamagnetic colloids under \textit{balanced}
\textit{triaxial magnetic fields }(BTMF)\textit{\ }- rotating
magnetic fields
spinning on a cone with the magic angle $\theta_{m}\approx54,7^{\circ}%
$\cite{Martin1,Martin2}. The emerging effective interaction was
similar to the London-Van der Waals force \cite{Osterman},
however, the structures that formed were far more intricate
\cite{Martin1,Martin2}.
 Here we develop the general theoretical framework for understanding this
important novel interaction, which is attracting a growing
interest in the self-assembly communities
\cite{NatureCommentMartin}. We explore how the scFI's
intrinsically strong many-body interactions drive the formation of
chains and membranes in systems of polarizable colloidal
particles\cite{Martin1,Martin2,Osterman} and how complex membrane
pair interactions give rise to novel swelling colloidal foam
states. The scFI generates complex structures (chains, membranes
and foams) while the, Van der Waals-like, \textit{incoherent
fluctuation interaction} (icFI) - only forms phase-separated
lumps, or droplets, of matter within a two phase system \cite{Van
Der Waals}. We discuss why these two related forces result in such
different structures.

Consider dipolar magnetic (dielectric) particles such as colloidal
beads without permanent moments. Here we use magnetic notation but
all results apply to the electric formalism (paramagnets become
dielectrics). Beads with index $i$, are placed in a spatially and
temporally fluctuating magnetic (electric) field
$\mathbf{B}_{0,i}=\mu_{0}\mathbf{H}_{0,i}(t)$ (with $\mu_{0}$ the
vacuum permeability). The field changes on a typical timescale
$\tau_{H}$ fulfilling the condition
$\tau_{M}\ll\tau_{H}\ll\tau_{visc}$ with $\tau_{M}$ the dipolar
relaxation time and $\tau_{visc}$ the characteristic time for the
bead's motion in the surrounding viscous fluid. Under these
conditions the beads' magnetizations $\mathbf{M}_{i}$ are
equilibrated, while their position coordinates $\mathbf{R}_{i}$
respond much more slowly and feel a net
time-averaged force. The free energy functional $\mathcal{F}(\mathbf{M}%
_{i};\mathbf{H}_{i,0})$ for $N$ interacting beads with volume
$V_{b}$ is then given by \cite{Landau-Lifshitz}: $\left(
\mu_{0}V_{b}\right)
^{-1}\mathcal{F}=-\frac{1}{2}\sum_{i=1}^{N}\mathbf{M}_{i}\mathbf{H}_{0,i}$.
Note, that due to the large moments
$\mathbf{m}_{i}=\mathbf{M}_{i}V_{i}$ of the beads with diameters
$D>1$ $\mu m$ we have $\mathcal{F\gg}$ $k_{B}T$ so that additional
contributions of the configurational entropy can be safely
neglected. The magnetization
$\mathbf{M}_{i}=\hat{\chi}_{b,i}\mathbf{H}_{i,loc}$ is given by
the \textit{total local} field $\mathbf{H}_{i,loc}$ and the
\textit{bead susceptibility} tensor
$\hat{\chi}_{b,i}=(1+\hat{L}_{i}\chi)^{-1}\chi$ with $\chi$ the
beads's \textit{material susceptibility} and $\hat{L}_{i}$ its
demagnetization tensor \cite{Landau-Lifshitz}. The dipolar
interaction between the $i$-th and $j$-th bead is given by the
dipole-dipole coupling tensor
$\hat{T}_{ij}=\varphi_{ij}\hat{t}(\mathbf{b}_{ij})$, with $\varphi_{ij}%
\equiv\varphi(\mathbf{R}_{i},\mathbf{R}_{j})=V_{b}/4\pi\left\vert
\mathbf{R}_{ij}\right\vert ^{3},$ $\mathbf{R}_{ij}\equiv\mathbf{R}%
_{i}-\mathbf{R}_{j}$ and the tensor $\hat{t}(\mathbf{b}_{ij})=\hat
{1}-3\mathbf{b}_{ij}\otimes\mathbf{b}_{ij}$ (dyadic product) with the bonding
unit vector $\mathbf{b}_{ij}=\mathbf{R}_{ij}/\left\vert \mathbf{R}%
_{ij}\right\vert $. The local field is a superposition of external and all
dipole induced fields and can be written as $\mathbf{H}_{i,loc}=\hat{\chi
}_{b,i}^{-1}\sum_{j}\hat{\chi}_{eff,ij}\mathbf{H}_{j,0}$ in terms of the
\textit{effective susceptibility }tensor\textit{ }$\hat{\chi}_{eff,\text{ }%
ij}=$\ $\left(  \hat{\chi}_{b,i}(\hat{1}+\hat{\chi}_{b,j}\hat{T})^{-1}\right)
_{ij}$ - a $3\times3$ matrix for each $i,j$ -see details in \cite{Kulic2}.

In the following we study a \textit{spatially coherent excitation} with the
property $\overline{H_{i,0}^{\alpha}H_{j,0}^{\beta}}\equiv C^{\alpha\beta
}=\delta_{\alpha\beta}H_{0}^{2}$ ($i,j=1,..,N;$ $\alpha,\beta=x,y,z$), where
bar means averaging over time or over random time-dependent fields. Note that
the correlation function $C_{\alpha\beta}$ comprises also the special case of
BTMF - studied numerically and experimentally in
\cite{Martin1,Martin2,Osterman}, where the field rotates along the z-axis with
the frequency $\omega$, i.e. $\mathbf{H}_{0}=H_{0}(\sqrt{2}\cos\omega
t,\sqrt{2}\sin\omega t,1)$\cite{NoteNonBallancedTMF}. After averaging of
$\mathcal{F}$ over $H_{i,0}^{\alpha}$ in all directions $\alpha=x,y,z$ (see
details in \cite{Kulic2}) one obtains an elegant expression for the scFI-free-energy%

\begin{equation}
\mathcal{\bar{F}}_{scFI}\left(  \mathbf{H}_{0},\{\mathbf{R}_{i}\}\right)
=-\frac{\mu_{0}}{2}H_{0}^{2}V_{b}\sum_{i,j}Tr\left\{  \hat{\chi}%
_{eff,ij}\right\}  . \label{chi-eff}%
\end{equation}
in terms of $Tr\left\{  \hat{\chi}_{eff,ij}\right\}  =\chi_{eff,ij}^{xx}%
+\chi_{eff,ij}^{yy}+\chi_{eff,ij}^{zz}$ - the trace of $\hat{\chi}_{eff,ij}.$
As $\hat{\chi}_{eff,ij}$ describes the effective coupling of the $i$-th and
$j$-th bead \textit{in presence of all other beads}, it gives rise to
many-body effects. These are crucial for the hierarchical structuring of
colloids, as confirmed in numerical simulations \cite{Martin1,Martin2} and
experiments \cite{Martin1,Martin2,Osterman}, where dimers, molecules, chains,
branched chains, membranes and foams are formed, instead of droplets and
close-packed 3D crystal structures. The elegant trace-formula Eq.\ref{chi-eff}
forms the basis for our further study of scFI systems. Note here that, in
contrast to scFI - described by Eq.\ref{chi-eff}, in standard incoherently
excited FI \ (icFI) systems, with $\overline{H_{i,0\alpha}H_{j,0\beta}}\equiv
C_{\alpha\beta}^{ij}=\delta_{ij}\delta_{\alpha\beta}H_{0}^{2}$ (note the
Kronecker symbol $\delta_{ij}$), the summation of $Tr\left\{  \hat{\chi
}_{eff,ij}\right\}  $ in $\mathcal{\bar{F}}$ includes only $i=j$ terms, giving
rise to the usual Van der Waals(VdW)-like forces, preferring the formation of
simple droplets only \cite{Van Der Waals}. In the following, we study first
the bead-bead interaction, then the formation of chains and membranes, and
finally the assembly of membranes in the form of foams in scFI systems.

\textit{Dimer formation.} In a first step, let us consider a very
dilute system. Here the \textit{pairwise} bead-bead interactions
should (at first glance) dominate in $\mathcal{\bar{F}}_{scFI}$.
For two spherical beads with indices $i,j$, we have
$\hat{L}_{\alpha\beta}=(1/3)\delta_{\alpha\beta}$,
$(\hat{\chi}_{b})_{\alpha\beta}=\chi_{b}\delta_{\alpha\beta}$
(with $\chi
_{b}\leq3$) and Eq.(\ref{chi-eff}) gives $Tr\left\{  \hat{\chi}_{eff,ij}%
^{(b)}\right\}  =3\chi_{b}(1-\chi_{b}\varphi_{ij})/(1+\chi_{b}\varphi
_{ij})(1-2\chi_{b}\varphi_{ij})$. This means that the bead-bead interaction is
attractive for\textit{\ }$\chi_{b}>0$. For large distances $\left\vert
\mathbf{R}_{ij}\right\vert \gg V_{b}^{1/3}$, the interaction is a power-law
$\mathcal{\bar{F}}_{ij}^{(bb)}=-(3/8\pi^{2})(V_{b}\chi_{b})^{3}\mu_{0}%
H_{0}^{2}\left\vert \mathbf{R}_{ij}\right\vert ^{-6}$ giving rise to an
isotropic and short-range, attractive VdW-like force $\mathbf{F}_{\mathbf{ij}%
}\propto\left\vert \mathbf{R}_{ij}\right\vert ^{-7}$. This interesting
two-body result was first obtained in the seminal papers
\cite{Martin1,Martin2} and confirmed experimentally \cite{Osterman}.
\begin{figure}[pt]
\begin{center}
\includegraphics[
width=3.2in]{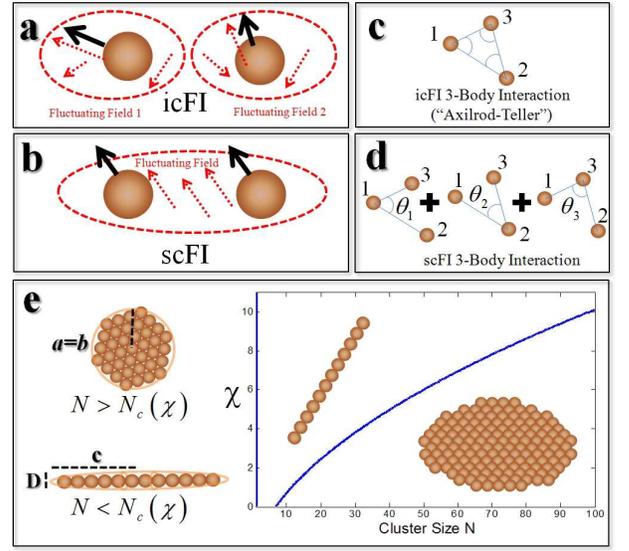}
\end{center}
\caption{a) The incoherent (Van der Waals-like) interaction icFI
and b) the spatially coherent fluctuation interaction scFI are
both induced by field fluctuations - but with different spatial
correlations. c),d) While the 2-body forces are similar in both
cases, the 3-body forces have different angular character and are
longer ranged for scFI, Eq.(\ref{3-body}). e) The phase diagram of
clusters of size $N$ and material susceptibility $\chi$. The
colloidal clusters are growing over time \cite{Martin1,Martin2}
and undergoing a transition from linear chains to membranes beyond
a critical size
$N_{c}\left(  \chi\right)  $. }%
\end{figure}

\textit{Many-body force}. At first glance, the isotropic VdW-like two-body
interaction $\propto\left\vert \mathbf{R}_{ij}\right\vert ^{-6}$ appears to
favor droplet-like, bulk structures. However, this contradicts numerical
simulations \cite{Martin1,Martin2} and experiments
\cite{Martin1,Martin2,Osterman}, which show a clear tendency for chain and
membrane formation. What is the microscopic origin of these complex structures
in scFI systems? The reason hides in the specificity and the strength of the
many-body interactions, which mark the real difference between scFI and icFI
systems. In a typical icFI system, the 3-body interaction is described by an
Axilrod-Teller-like 3-body potential $\mathcal{\bar{F}}_{A.-T.}^{(ijk)}%
=A_{ijk}\left\vert \mathbf{R}_{ij}\right\vert ^{-3}\left\vert \mathbf{R}%
_{jk}\right\vert ^{-3}\left\vert \mathbf{R}_{ki}\right\vert ^{-3}$, with
$A_{ijk}\propto\chi_{b}^{3}\left(  1+\cos\theta_{i}\cos\theta_{j}\cos
\theta_{k}\right)  $ \cite{Teller}. Due to its weaker ($\sim\chi_{b}^{3}%
R^{-9}$) scaling, it is typically small and overridden by the $2$-body
VdW-like interaction $\sim\chi_{b}^{2}R^{-6}$ \cite{NOTEAxilrodTeller}, giving
rise to close packed droplets in icFI systems. In sharp contrast, in scFI
systems the 3-body interaction $\mathcal{\bar{F}}_{scFI}^{(ijk)}$ scales very
differently with distance, as seen from a $O\left(  \chi_{b}^{3}\varphi
_{ij}^{3}\right)  $ expansion of Eq.(\ref{chi-eff}) - see details in
\cite{Kulic2}, which gives
\begin{equation}
\mathcal{\bar{F}}_{scFI}^{(ijk)}=-\beta\sum\nolimits_{i,j,k}^{\prime}%
\frac{3\cos^{2}\theta_{k}-1}{\left\vert \mathbf{R}_{ik}\right\vert
^{3}\left\vert \mathbf{R}_{kj}\right\vert ^{3}}\label{3-body}%
\end{equation}
with $\beta=(3/64\pi^{2})\mu_{0}H_{0}^{2}\chi_{b}^{3}V_{b}^{3}$ and the sum
running over all $k\neq i,j$ (for angles $\theta_{k}$ cf. Fig.1d). Remarkably,
$\mathcal{\bar{F}}_{scFI}^{(ijk)}$ is of the same order as the $2$-body
interaction $\mathcal{\bar{F}}_{scFI}^{(ij)}\sim-\chi_{b}^{3}\left\vert
\mathbf{R}_{ij}\right\vert ^{-6}$. In fact, the $2$-body interaction is
formally \textit{contained }in $\mathcal{\bar{F}}_{scFI}^{(ijk)}$ (for $k\neq
i=j$) showing us the pitfall in the dimer formation section: for $scFI$ the
$2$-body interactions are physically\textit{\ inseparable} from the $3$-body
ones - at any $\chi_{b}$. Interestingly, $\mathcal{\bar{F}}_{scFI}^{(ijk)}$
has a simple angular dependence explaining scFI's tendency to destroy 3D bulk
structures: the $-\cos^{2}\theta_{k}$ term favors $\theta_{k}=0$ or $\pi\ $-
i.e. the colloidal chains found in the experiments
\cite{Martin1,Martin2,Osterman}.

\textit{Chains and membranes.} The $-\cos^{2}\theta_{k}$ term in
the $3$-body force of Eq.\ref{3-body} explains the initial
formation of chains. To capture
quantitatively their transition to membranes, higher $O\left(  \chi_{b}%
^{3}\varphi_{ij}^{3}\right)  $ terms beyond Eq.\ref{3-body} are necessary. It
is however conceptually more instructive to take a more macroscopic approach,
where dense chains/membranes are modelled by prolate/oblate ellipsoids. Here,
$\hat{\chi}_{eff,ij}$ in Eq.\ref{chi-eff} is replaced by its shape-dependent
continuum limit $\hat{\chi}^{(L)}=\chi(1+\hat{L}\chi)^{-1}$, where $\hat{L}$
is now the demagnetization tensor of the composite ellipsoidal structure with
dimensions $a=b\neq c$ and volume $V=(4\pi/3)a^{2}c$ \cite{Landau-Lifshitz}.
Here, $\chi$ is the \textit{composite material susceptibility,} which is due
to local field effects in the bead aggregate. In dense systems like chain and
membranes one has $\chi>\chi_{b}$. The free-energy $\mathcal{\bar{F}}_{self}$
is in this case given by%
\begin{equation}
\mathcal{\bar{F}}_{self}(\mathbf{H}_{0},\hat{L})=-\frac{1}{2}\mu_{0}H_{0}%
^{2}VTr\left\{  \hat{\chi}^{(L)}\right\}  . \label{F-macro1}%
\end{equation}

In the continuum limit, when the number of beads is large
($N\gg1$), chains become\textit{\ extreme prolate ellipsoids} with
a long semi-axis $c\sim ND$, a short semi-axis$\ a\sim D$ and the
demagnetization factors $L_{a}=L_{b}\gg L_{c}\equiv L^{ch}\sim
N^{-2}\ln N$. Membranes, on the other hand, can be seen\textit{\
}as\textit{\ extreme oblate ellipsoids }with two identical
half-axes $a=b\sim\sqrt{N}D$ $\ ,\ c\sim D$ and $L_{c}\gg
L_{a}=L_{b}\equiv L^{m}\sim N^{-1/2},$cf. Fig 1e. In both cases,
for chains and membranes, $\mathcal{\bar{F}}_{self}$ is dominated
by the smallest demagnetization factors $L^{m/ch}\rightarrow0$.
Since for membranes two demagnetization factors vanish, while for
chains only one vanishes, the energy of a membrane is always
smaller than that of the chain for $N\rightarrow\infty$, i.e.
$\mathcal{\bar{F}}^{m}<\mathcal{\bar{F}}^{ch}$. For finite $N$,
the \textit{chain-membrane transition }is reached on the line
$N=N_{c}(\chi)$, where
$\mathcal{\bar{F}}^{m}(N_{c})=\mathcal{\bar{F}}^{ch}(N_{c})$. As
seen from Fig.1e, $N_{c}$ grows with the material susceptibility
$\chi$. For $\chi\approx1-3$, we estimate $N_{c}\approx10-20$ -
close to the experimental value $N_{c}\approx10$ for the
chain-membrane transition \cite{Osterman}. The agreement suggests
that the continuum approach quantitatively captures the behavior
for finite $N$.

\textit{Interaction of membranes.} Once they emerge, what is the fate of the
membranes as they continue growing? How do they interact and mutually order
during growth? To answer these questions, we need the\textit{\ 2-membrane
interaction} for arbitrary membrane orientations $\mathbf{n}_{1,2}$ and
anisotropic susceptibilities (of thin oblates) $\hat{\chi}_{i}^{(L)}%
=\chi(1+\hat{L}_{i}\chi)^{-1}$. For large distances ($\varphi_{12}%
\ll1,\left\vert \mathbf{R}_{12}\right\vert \gg V_{m}^{1/3}$) one expands
$\hat{\chi}_{eff,12}\approx\hat{\chi}_{1}^{(L)}(1-\varphi_{12}$ $\hat
{t}(\mathbf{b}_{12})\hat{\chi}_{\mathbf{2}}^{(L)})$ and the long range
interaction energy in Eq.\ref{chi-eff} reads for identical membranes (see
details in \cite{Kulic2}),%

\begin{equation}
\mathcal{\bar{F}}_{int}=\alpha\frac{C_{1}^{2}+C_{2}^{2}+\frac{1-\gamma}%
{3}C_{3}^{2}-(1-\gamma)C_{1}C_{2}C_{3}-\frac{2}{3}}{\left\vert \mathbf{R}%
_{12}\right\vert ^{3}}\label{me-me}%
\end{equation}
with $\alpha=3(1-\gamma)\chi_{\max}^{2}\mu_{0}H_{0}^{2}V_{m}^{2}/16\pi,$ and
$\gamma=\chi_{\min}/\chi_{\max}$ the ratio of the minimal/maximal eigenvalue
of the membrane susceptibility tensor $\hat{\chi}^{(L)}$. The dimensionless
factors $C_{1}=\mathbf{n}_{1}\cdot\mathbf{b}_{12}$, $C_{2}=\mathbf{n}_{2}%
\cdot\mathbf{b}_{12}$, $C_{3}=\mathbf{n}_{1}\cdot\mathbf{n}_{2}$ reveal all
the geometrical beauty of scFI: the 2-membrane interaction is angle dependent
and repulsive in many configurations - see Fig.2a. Notably, for fixed
$\left\vert \mathbf{R}_{12}\right\vert $, $\mathcal{\bar{F}}_{int}$ becomes
minimal for the orthogonally \textit{twisted} membrane orientation with
$\mathbf{n}_{1}\perp\mathbf{n}_{2}$, $\mathbf{n}_{1}\perp\mathbf{b}_{12}$ and
$\mathbf{n}_{2}\perp\mathbf{b}_{12}$ ($C_{1/2/3}=0$). The twisted membranes
attract each other since $\mathcal{\bar{F}}_{int}^{\left(  tw\right)  }<0$ (up
to the point of mutual contact), as in the \textit{coplanar} case, yet the
\textit{twisted} configuration has lower energy. This interesting result
should affect the kinetics of membrane formation: If two distant membranes
start growing within a large distance they will rotate to a $90^{\circ}$
position before touching. Therefore, some type of glassy state in their
orientation may be kinetically favored. In other relevant configurations, such
as the \textit{top}, with two out of plane parallel membranes ($C_{1/2/3}=1$)
or the \textit{generic} one (cf. Fig.2a), the interaction is repulsive with
$0<\mathcal{\bar{F}}_{int}^{(gen)}<\mathcal{\bar{F}}_{int}^{(top)}$.
\begin{figure}[pt]
\begin{center}
\includegraphics[
width=3.2in]{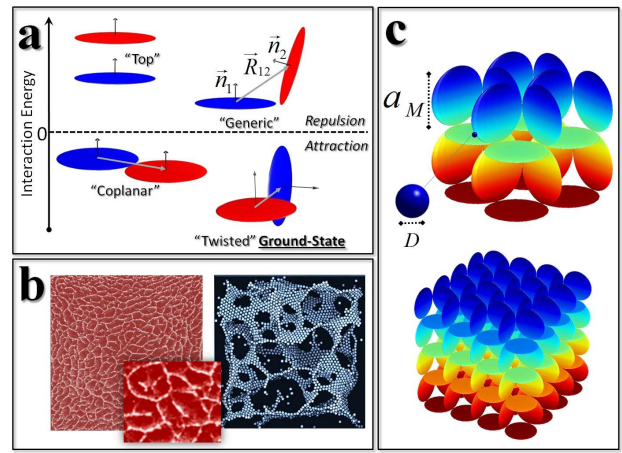}
\end{center}
\caption{(a)The scFI is unexpectedly complex: The 2-membrane
interaction is attractive or repulsive depending on orientation,
cf. Eq.\ref{me-me}, with a ground state in the "twisted"
configuration. (b) The large scale structure of a dipolar foam
(from \cite{Martin1,Martin2}) in experiment (left) and in
simulation (right). (c) The theoretical 3D shelf-model for the
foam structure.}%
\end{figure}

\textit{Emergence of foams.} Simulations and experiments
\cite{Martin1,Martin2} provide empirical evidence for a hollow foam-like
superstructure (cf. Fig 2b). What is the physical mechanism behind such
dipolar foam formation? We have seen above that large aggregates prefer to
form membranes, and that these membranes mutually interact. Specifically, when
two distant membranes are stacked over each other, they repel ($\mathcal{\bar
{F}}_{int}^{(top)}>0$). In the opposite limit (contact distance), a simple
estimate implies their preference to split as well \cite{NOTE2Membranes}. It
is this remarkable reluctance of membranes to mutually stack that in fact sets
the microscopic structure of the foam: It is formed out of the thinnest
possible membrane patches, whose thickness is collapsed onto the smallest
available physical scale - the bead size $D$. The characteristic lateral size
$a_{M}$ of these membrane patches, on the other hand, is set by the bead
volume fraction in the container $f_{V}=V_{b}^{tot}/V\ll1$ (with $V_{b}^{tot}$
the\ total volume of all beads and $V$ the container's volume). By assuming a
cubic shelf structure as an ansatz, cf. Fig 2c, one obtains a patch size
$a_{M}=3D/f_{V}$.

\textit{Pressure in scFI systems. }Since the membranes tend to
grow, yet repel on the average, the container walls will feel a
positive net pressure $p=-\partial\mathcal{\bar{F}}/\partial V>0$.
The total free energy
$\mathcal{\bar{F}=\bar{F}}_{self}+\mathcal{\bar{F}}_{int}$ has
positive contributions from both the membrane self-energy $\mathcal{\bar{F}%
}_{self}$ and their pairwise interactions $\mathcal{\bar{F}}_{int}$. We can
qualitatively mimic the foam structure \cite{Martin1,Martin2} using the simple
shelf-like cubic lattice ansatz with the mesh-size $a_{M}$ (Fig.2c). The
self-energies of finite membranes depend on their demagnetization factors
$L^{m}$, which for the assumed simple shelf-like cubic lattice are given by
$f_{V}$ as $L^{m}\approx\alpha_{m}f_{V}$ for $f_{V}\ll1$ with $\alpha
_{m}\approx1/4$. For $\chi\gg1$ and $L^{m}\chi\ll1$ one has $\mathcal{\bar{F}%
}_{self}\approx0.5\chi^{2}f_{V}p_{0}V_{b}^{tot}$ (from Eq.\ref{F-macro1}) with
$p_{0}=\mu_{0}H_{0}^{2}/2$. The interaction energy $\mathcal{\bar{F}}%
_{int}=(S/8\pi)\chi^{2}f_{V}p_{0}V_{b}^{tot}$ is calculated by
explicitly summing over all interactions (given by Eq.\ref{me-me})
of a membrane with other membranes in the cubic shelf lattice,
with a numeric constant $S$ $\approx10$ calculated by numerical
lattice summation \cite{Kulic2}. As a result the foam's
pressure is%
\begin{equation}
p\approx\frac{1}{2}\mu_{0}\chi^{2}f_{V}^{2}H_{0}^{2}.\label{pressure}%
\end{equation}
The foam's pressure can assume notable magnitudes. For moderate volume
fractions, fields and susceptibilities ($f_{V}\approx5\cdot10^{-2},$ $\mu
_{0}H_{0}\approx20mT,$ and $\chi\approx10$ in densely packed $Ni$-beads
membranes) we obtain $p\approx40$ $Pa$. Since $p\propto H_{0}^{2}$, the
pressure is very sensitive to the strength of excitation $H_{0}$ and can lead
to remarkably strong swelling of the foam against gravity. The latter effect
can be used to experimentally test Eq.\ref{pressure}. The equilibrium foam
height $h$ is reached once the internal and the gravitational pressure
balance, i.e. $p\approx\Delta\rho gf_{V}h$. For the density contrast
$\Delta\rho\simeq8\cdot10^{3}kg/m^{3}$ of water immersed $Ni$-beads
\cite{Martin1,Martin2}, $g\simeq10m/s^{2},$ the foam will swell strongly up to
$h\sim1$ $cm$. More precisely, due to gravity, $f_{V}$ and $p$ vary with $h$,
i.e. $p(f_{V})=p_{\max}(1-h/2h^{0})^{2}$ giving the maximal height $h_{\max
}=2h^{0}\sim2$ $cm$.

\textit{Conclusion.} We have studied the formation of hierarchical
superstructures in dipolar particle systems driven by the spatially coherent
fluctuation interaction. The pronounced many-body interactions give rise to a
growth of anisotropic assemblies - chains, then membranes once a critical
cluster size is reached. In a container of finite size, smaller membrane
patches are formed, which, contrary to the case of attracting beads, repel on
average, thus giving rise to dipolar foam structures. The foam exerts a
positive pressure on the walls of the container due to the tendency of
membranes to increase their surface areas as well as their mutual repulsion.
The dipolar foam represents a new and intriguing state of colloidal matter,
formed by a delicate interplay of an attractive local interaction and a net
repulsive longer range force. Remarkably, both types of forces are born out of
a single, conceptually simple interaction - the scFI (Eq.\ref{chi-eff}). Being
so simple to generate, yet rich and intricate in its effects, makes it a
promising tool for future applications in self-assembly and nano-science.

\textit{Acknowledgements. }We thank A.Johner, H.Mohrbach, M.Greenall for
discussions and comments.


\begin{thebibliography}{99}                                                                                               %


\bibitem {ColdAtoms}V. I. Balykin, V. G. Minogin and V. S. Letokhov, Rep.
Prog. Phys. 63, 1429 (2000).

\bibitem {Colloidal Suspensions}W.B.Russel, D.A.Saville, and W.R. Schowalter,
Colloidal Dispersions, Cambridge, Cambridge University Press (1989)

\bibitem {Levitron}Berry, M V, \ Proc.Roy.Soc.Lond. A 452, 1207, (1996)

\bibitem {Block-Copolymers}I. W. Hamley, The Physics of Block Copolymers ,
Oxford Science Publications, (1999)

\bibitem {Labyrinthine}A. Tsebers, M. Maiorov. Magnetohydrodynamics, vol. 16,
21--27 (1980); Rosensweig R E, Zahn M and Shumovich R J. Magn. Magn. Mater.
39, 127-32 (1983)

\bibitem {Nuclear-Pasta}D. G. Ravenhall, C. J. Pethick, and J. R. Wilson,
Phys. Rev.Lett. 50, 2066 (1983); M. Hashimoto, H. Seki, and M. Yamada, Prog.
Theor. Phys. 71, 320 (1984).

\bibitem {Martin1}J. E. Martin, R. A. Anderson, R. L. Williamson, J. Chem.
Phys. \textbf{118}, 1557 (2003)

\bibitem {Martin2}J. E. Martin, E. Venturini, G. L. Gulley, J. Williamson,
Phys. Rev. \textbf{E 69}, 021508-1 (2004)

\bibitem {Osterman}N. Osterman, I. Poberaj, J. Donikar, D. Frenkel, P.Ziherl,
D. Babi\'{c}, Phys. Rev. Lett. \textbf{103}, 228301 (2009)

\bibitem {NatureCommentMartin}J. F. Douglas, Nature 463, 302 (2010)

\bibitem {Van Der Waals}V.A. Parsegian, Van Der Waals Forces, Cambridge
University Press (2006)


\bibitem {Landau-Lifshitz}L. D. Landau, E. M. Lifshitz, Electrodyanamics of
Continuous Media, Oxford: Pergamon Press (1989)

\bibitem {Kulic2} Supplementary information material.

\bibitem {NoteNonBallancedTMF}Note that in the most general case found in
literature \cite{Martin1},\cite{Martin2} the triaxial field is unballanced and
can have an arbitrary in-plane $H_{\parallel}$ and perpendicual magnitude
$H_{\perp}:$ $\mathbf{H}_{0}=(H_{\parallel}\cos\omega t,H_{\parallel}%
\sin\omega t,H_{\perp})$ with $2H_{\parallel}^{2}+H_{\perp}^{2}=H_{0}^{2}.$
One can show that in this case the interaction can be linearly decomposed into
a ballanced triaxial field (BTMF) magic angle interaction and a residual
dipole-dipole interaction along the orthogonal direction\cite{Kulic2}. The
latter is well understood while the former is new and investigated here.

\bibitem {Teller}B. M. Axilrod, E. Teller, J. Chem. Phys. \textbf{11}, 299
(1943); Yu. S. Barash, V. L. Ginzburg, Sov. Phys. Uspekhi, \textbf{143}, 345 (1984)

\bibitem {NOTEAxilrodTeller}The $\chi^{2}$ scaling is valid for equilibrium
conditions, where a detailed ballance between the moments and (thermal or
quantum) bath hold. The scaling for icFI\ switches to $\propto\chi^{3}$ if the
fluctuating field is exogeneous (externally set) and non-equilibrium as in the
present scFI case \cite{Kulic2}.

\bibitem {NOTE2Membranes}When a thick membrane (thickness $2D$, radius $R$ and
volume $2V_{m}$) is cut into two parallel membranes (thickness $D$ and radius
$R$ each) and separated to infinite distance there is a gain in the energy
$\Delta\mathcal{F}=2\mathcal{F}_{1m}-\mathcal{F}_{2m}\approx-V_{2m}L\chi
^{2}(1-(1+\chi)^{-2})<0$ for $L\chi\ll1$, where $L\propto D^{3/2}V_{2m}%
^{-1/2}$. Physically, the second membrane lying above the first one is
repelled to increase the local fields with respect to the thicker membrane
case \cite{Kulic2}.


\end{thebibliography}
\end{document}


\title{Coherent Fluctuation Interaction}
\author{Igor M. Kuli\'{c} and Miodrag L. Kuli\'{c}}
\date{\today }
\maketitle

\part{Supplementary Information}

In this supplement we derive the key relations used in the main text. It turns
out that the \textit{spatially coherent fluctuation interaction} (scFI) as
given by Eq.1 often permits rather compact and elegant calculations which
directly lead to physical insights.

\section{Derivation of the "Trace-Formula" for scFI (Eq. 1)}

Here we first derive the central "Trace-Formula" (Eq.1) from the main text.
Starting from the basic expression for the free-energy $\mathcal{F}%
\{\mathbf{M}_{i},\mathbf{H}_{i,0}\}$ in an inhomogeneous external field
$\mathbf{H}_{i,0}$ (with the bead-volume $V_{b},$ magnetization $\mathbf{M}%
_{i}$) given by:
\begin{equation}
\left(  \mu_{0}V_{b}\right)  ^{-1}\mathcal{F}=\sum_{i=1}^{N}\left(  \frac
{1}{2}\mathbf{M}_{i}\chi_{b}^{-1}\mathbf{M}_{i}-\mathbf{M}_{i}\mathbf{H}%
_{i,0}\right)  +\frac{1}{2}\sum_{i,j\mathbf{\neq}i}\mathbf{M}_{i}\hat{T}%
_{ij}\mathbf{M}_{j} \label{S1}%
\end{equation}
where the summation goes over $N$ beads, $\mu_{0}$ is the vacuum permeability
and $\chi_{b}$ (we assume equal beads) is the \textit{bead susceptibility} in
external field. The dipole-dipole interaction is described by $\hat{T}%
_{ij}=\varphi_{ij}\hat{t}(\mathbf{b}_{ij})$ ( $i\neq j$ ), with $\varphi
_{ij}\equiv\varphi(\mathbf{R}_{i},\mathbf{R}_{j})=V_{b}/4\pi\left\vert
\mathbf{R}_{ij}\right\vert ^{3},$ $\mathbf{R}_{ij}\equiv\mathbf{R}%
_{i}-\mathbf{R}_{j}\neq0$ and the tensor $\hat{t}(\mathbf{b}_{ij})=\hat
{1}-3\left\vert \mathbf{b}_{ij}\right\rangle \left\langle \mathbf{b}%
_{ij}\right\vert $ (dyadic product in bra-ket form) with the "bonding" unit
vector $\mathbf{b}_{ij}=\mathbf{R}_{ij}/\left\vert \mathbf{R}_{ij}\right\vert
$.

After minimization w.r.t. $\mathbf{M}_{i}$ we obtain the equilibrium
free-energy (for fixed coordinates $\{\mathbf{R}_{i}\}$)
\begin{equation}
\mathcal{F}_{scDI}\{\mathbf{H}_{i,0}\}=-\frac{1}{2}\mu_{0}V_{b}\sum
_{i}\mathbf{M}_{i}\mathbf{H}_{i,0}.\label{S2}%
\end{equation}
The magnetizations $\mathbf{M}_{i}=\chi_{b}\mathbf{H}_{i,loc}$ are given by
the \textit{total local} fields $\mathbf{H}_{i,loc}$ , i.e. by the
superposition of external and all dipole induced fields. These local fields
are related to the imposed external fields $\mathbf{H}_{i,0}$ via the matrix
(in continuum limit integral) equation%

\begin{equation}
\sum_{j}\left(  \delta_{ij}+\chi_{b}\hat{T}_{ij}\right)  \mathbf{H}%
_{j,loc}=\mathbf{H}_{i,0}.\label{S3}%
\end{equation}
The last line is strictly valid if we adopt the practical convention that
$\hat{T}_{ii}=0$ for two identical particle indices.The formal solution for
the local fields $\mathbf{H}_{i,loc}$ gives $\chi_{b}\mathbf{H}_{i,loc}%
=\sum_{j}\hat{\chi}_{eff,ij}\mathbf{H}_{i,0}$ with the operator $\hat{\chi
}_{eff}=\chi_{b}(\hat{1}+\chi_{b}\hat{T})^{-1}$. Its components can be
formally written as $\hat{\chi}_{eff,ij}\equiv\left\langle \mathbf{R}%
_{i}\right\vert \chi_{b}(\hat{1}+\chi_{b}\hat{T})^{-1}\left\vert
\mathbf{R}_{j}\right\rangle $ where $\left\langle \mathbf{R}_{i}\right\vert
\hat{A}\left\vert \mathbf{R}_{j}\right\rangle =A_{ij}$ stands for the
components of the operator $\hat{A}$ with respect to particles $i$ and $j.$
Note that each component $\hat{\chi}_{eff,ij}$ , for fixed $i$ and $j,$ is
itself a 3 dimensional 2-tensor (a $3\times3$ matrix in 3 dimensional space).
Note also, that the $i,j$ components are to be evaluated \textit{after} the
operator inversion, which is at the very origin of many-body forces (cf. below).

Now, the free-energy is given by $\mathcal{F}\left(  \mathbf{H}_{0}%
,\{\mathbf{R}_{i}\}\right)  $. Since $\mathbf{H}_{i,0}$ is a time (and
possibly position) fluctuating field, the average $\mathcal{\bar{F}}$ over
fluctuations gives the average free-energy
\begin{align}
\mathcal{\bar{F}}\left(  \mathbf{H}_{0},\{\mathbf{R}_{i}\})\right)   &
=-\frac{\mu_{0}}{2}V_{b}\sum_{i,j}\overline{\mathbf{H}_{i,0}\hat{\chi
}_{eff,ij}\mathbf{H}_{j,0}}\label{S4}\\
&  =-\frac{\mu_{0}}{2}V_{b}\sum_{i,j,\alpha,\beta}C_{ij}^{\alpha\beta}%
\chi_{eff,ij}^{\alpha\beta},\nonumber
\end{align}
where the field correlation function $C_{ij}^{\alpha\beta}=\overline
{H_{i,0}^{\alpha}H_{j,0}^{\beta}}$ and $\alpha,\beta=x,y,z$ and $i,j=1,2....N$.

Two limiting cases for $C_{ij}^{\alpha\beta}$ are of interest:

(1) The \textit{spatially coherent fluctuation interaction} (scFI) with
$C_{ij}^{\alpha\beta}=C^{\alpha\beta}=\delta_{\alpha\beta}H_{0}^{2}%
+h^{\alpha\beta}$. The first term in $C^{\alpha\beta}$ is the time-averaging
over a statistically \textit{isotropic}, spatially coherent (uniform)
excitation while the second term $h^{\alpha\beta}$ describes the anisotropic
contributions. The "isotropic" case (with $C^{\alpha\beta}=\delta_{\alpha
\beta}H_{0}^{2}$) is the main subject in the manuscript.The corresponding
free-energy is given by
\begin{equation}
\mathcal{\bar{F}}_{scFI}\left(  \mathbf{H}_{0},\{\mathbf{R}_{i}\}\right)
=-\frac{\mu_{0}}{2}H_{0}^{2}V_{b}\sum_{i,j}Tr\left(  \hat{\chi}_{eff,ij}%
\right)  .\label{S5}%
\end{equation}
where $Tr\left(  \hat{\chi}_{eff,ij}\right)  \equiv\chi_{eff,ij}^{xx}%
+\chi_{eff,ij}^{yy}+\chi_{eff,ij}^{zz}$.

Note, the "isotropic" correlation function $C_{\alpha\beta}=\delta
_{\alpha\beta}H_{0}^{2}$ comprises also the case of \textit{balanced triaxial
magnetic fields (}BTMF) - studied numerically and experimentally in
literature, where the magnetic field rotates with \ the frequency $\omega$ on
the cone with the magic angle $\theta_{m}(=\arccos(1/\sqrt{3})\approx
54,7^{\circ}$ (with $\cos\theta_{m}=\sin\theta_{m}$). In this case we
parametrize the field $\mathbf{H}_{0}(t)=\sqrt{3}H_{0}(\sqrt{2}\sin\theta
_{m}\cos\omega t,\sqrt{2}\sin\theta_{m}\sin\omega t,\cos\theta_{m})$. This
gives $\mathbf{H}_{0}(t)=H_{0}(\sqrt{2}\cos\omega t,\sqrt{2}\sin\omega t,1)$
and $\overline{H_{i,0}^{\alpha}H_{j,0}^{\beta}}=\delta_{\alpha\beta}H_{0}^{2}$
i.e. indeed an isotropic interaction.

(2) \textit{The spatially\ incoherently excited fields} (different at all
particle positions) - (icFI) systems, with $C_{ij}^{\alpha\beta}=\delta
_{ij}C^{\alpha\beta}$, which again contains isotropic and anisotropic terms.
Note the $\delta_{ij}$ term destroying correlations between excitations for
different particles. In the isotropic icFI case ($C_{ij}^{\alpha\beta}%
=\delta_{ij}\delta_{\alpha\beta}H_{0}^{2}$) we have then:%
\begin{equation}
\mathcal{\bar{F}}_{icFI}\left(  \mathbf{H}_{0},\{\mathbf{R}_{i}\}\right)
=-\frac{\mu_{0}}{2}H_{0}^{2}V_{b}\sum_{i}Tr\left(  \hat{\chi}_{eff,ii}\right)
.\label{S6}%
\end{equation}
Note the difference between the two cases (1) and (2). In the incoherent case
(ii) $\mathcal{\bar{F}}_{icFI}\left(  \mathbf{H}_{0},\{\mathbf{R}%
_{i}\}\right)  $ contains a summation over index $i$ only, and includes only
diagonal terms $i=j$, which is analogous to the usual Van der Waals(VdW) -like
forces. However in case (1) it is the double summation over $i$ and $j$ which
gives rise to all interesting physics, in particular the strong many body
effects characteristic of scFI.

\section{Derivation of the Three-Body Forces for scFI (Eq. 2)}

In the case of the \textit{spatially coherent fluctuation interaction} (scFI)
the effective free-energy is given by Eq.(1) in the paper%

\begin{equation}
\mathcal{\bar{F}}_{scDI}\left(  \mathbf{H}_{0},\{\mathbf{R}_{i}\}\right)
=-\frac{\mu_{0}}{2}H_{0}^{2}V_{b}\sum_{i,j}Tr\left(  \hat{\chi}_{eff,ij}%
\right)  . \label{T1}%
\end{equation}
where $Tr\left(  \hat{\chi}_{eff,ij}\right)  \equiv\chi_{eff,ij}^{xx}%
+\chi_{eff,ij}^{yy}+\chi_{eff,ij}^{zz}$ is the trace of the effective many
body susceptibility tensors $\hat{\chi}_{eff},_{ij}$. In the simplest case of
isotropic magnetic beads with isotropic (scalar) susceptibility\ (with respect
to the external field) $\chi_{b},$ one has $\hat{\chi}_{eff},_{ij}%
\equiv\left\langle \mathbf{R}_{i}\right\vert \hat{\chi}_{eff}\left\vert
\mathbf{R}_{j}\right\rangle =\chi_{b}\left\langle \mathbf{R}_{i}\right\vert
(\hat{1}+\chi_{b}\hat{T})^{-1}\left\vert \mathbf{R}_{j}\right\rangle $ and
$\hat{\chi}_{eff}=\chi_{b}(\hat{1}+\chi_{b}\hat{T})^{-1}$ can be expanded in
terms of powers of $\chi_{b}\hat{T}$
\begin{equation}
\hat{\chi}_{eff}=\chi_{b}\hat{1}-\chi_{b}^{2}\hat{T}+\chi_{b}^{3}\hat{T}%
^{2}-\chi_{b}^{4}\hat{T}^{3}+... \label{T2}%
\end{equation}

The dimensionless free-energy $f_{scDI}\equiv\mathcal{\bar{F}}_{scDI}%
(\mathbf{H}_{0},\{\mathbf{R}_{i}\})/(\mu_{0}V_{b}H_{0}^{2}/2)=f_{1}%
+f_{2}+f_{3}+...$ is given by%
\begin{equation}
f_{scDI}=\underset{f_{1}}{\underbrace{-\chi_{b}N_{b}}}+\underset{f_{2}%
}{\underbrace{\chi_{b}^{2}Tr\left(  \sum_{i\neq j}^{N}\hat{T}_{ij}\right)  }%
+}\underset{f_{3}}{\underbrace{\chi_{b}^{3}Tr\left(  -\sum_{m=1}^{N}%
\sum_{i\neq m}\sum_{j\neq m}\hat{T}_{im}\hat{T}_{mj}\right)  }}+O\left(
\chi_{b}^{4}\hat{T}^{3}\right)  \label{T3}%
\end{equation}

The first term $f_{1}=-N_{b}\chi_{b}$ describes the self-energy of
(noninteracting) beads, where $N_{b}$ is the number of beads. The second
pairwise term (i.e. the first order dipole-dipole-like interaction)
$f_{2}=\chi_{b}^{2}Tr\left(  \hat{T}_{ij}\right)  =0$ is strictly zero for
scFI as the operator $\hat{T}_{ij}$ is always traceless. This is easy to see,
since in any orthogonal basis $\left\{  \left\vert \mathbf{e}_{\alpha
}\right\rangle ,\alpha=x,y,z\right\}  $ one has
\begin{equation}
Tr\left(  \hat{T}_{ij}\right)  =\varphi_{ij}Tr(\hat{t}(\mathbf{b}%
_{ij}))=\varphi_{ij}\sum_{\alpha}\left\{  \left\langle \mathbf{e}_{\alpha
}\right\vert \left\vert \mathbf{e}_{\alpha}\right\rangle -3\left\langle
\mathbf{e}_{\alpha}\right\vert \left\vert \mathbf{b}_{ij}\right\rangle
\left\langle \mathbf{b}_{ij}\right\vert \left\vert \mathbf{e}_{\alpha
}\right\rangle \right\}  \propto3-3=0. \label{T4}%
\end{equation}

The third term $f_{3}$ is the first (lowest order) non-trivial one. It is very
interesting since it contains the three-body interactions too. It consist of
terms of the form $Tr\left(  \hat{T}_{im}\hat{T}_{mj}\right)  $ which can be
easily evaluated
\begin{align}
-Tr\left(  \hat{T}_{im}\hat{T}_{mj}\right)   &  =-\tfrac{V_{b}^{3}}{4^{2}%
\pi^{2}\left\vert \mathbf{R}_{i}-\mathbf{R}_{m}\right\vert ^{3}\left\vert
\mathbf{R}_{j}-\mathbf{R}_{m}\right\vert ^{3}}Tr\left[  \left(  \hat
{1}-3\left\vert \mathbf{b}_{im}\right\rangle \left\langle \mathbf{b}%
_{im}\right\vert \right)  \left(  \hat{1}-3\left\vert \mathbf{b}%
_{jm}\right\rangle \left\langle \mathbf{b}_{jm}\right\vert \right)  \right]
\label{T5}\\
&  =-\tfrac{V_{b}^{3}}{16\pi^{2}\left\vert \mathbf{R}_{i}-\mathbf{R}%
_{m}\right\vert ^{3}\left\vert \mathbf{R}_{j}-\mathbf{R}_{m}\right\vert ^{3}%
}Tr\left[  \left(  \hat{1}+9\left\vert \mathbf{b}_{im}\right\rangle
\left\langle \mathbf{b}_{im}\right\vert \left\vert \mathbf{b}_{jm}%
\right\rangle \left\langle \mathbf{b}_{jm}\right\vert -3\left(  \left\vert
\mathbf{b}_{im}\right\rangle \left\langle \mathbf{b}_{im}\right\vert
+\left\vert \mathbf{b}_{jm}\right\rangle \left\langle \mathbf{b}%
_{jm}\right\vert \right)  \right)  \right] \nonumber\\
&  =-\tfrac{V_{b}^{3}}{16\pi^{2}\left\vert \mathbf{R}_{i}-\mathbf{R}%
_{m}\right\vert ^{3}\left\vert \mathbf{R}_{j}-\mathbf{R}_{m}\right\vert ^{3}%
}\left(  3+9\left\langle \mathbf{b}_{im}\right\vert \left\vert \mathbf{b}%
_{jm}\right\rangle ^{2}-3\left(  1+1\right)  \right) \nonumber\\
&  =-\tfrac{3V_{b}^{3}}{16\pi^{2}\left\vert \mathbf{R}_{i}-\mathbf{R}%
_{m}\right\vert ^{3}\left\vert \mathbf{R}_{j}-\mathbf{R}_{m}\right\vert ^{3}%
}\left(  3\left\langle \mathbf{b}_{im}\right\vert \left\vert \mathbf{b}%
_{jm}\right\rangle ^{2}-1\right) \nonumber
\end{align}

The scalar product of the (unit) bonding vectors appearing above is nothing
else but the cosine of the angles between them as seen from the $m-th$
particle (cf. Fig 1d main text) i.e. $\left\langle \mathbf{b}_{im}\right\vert
\left\vert \mathbf{b}_{mj}\right\rangle =\cos\left(  \measuredangle
_{i,m,j}\right)  .$ Finally, the term $\mathcal{\bar{F}}_{3}$ reads%

\begin{align}
\mathcal{\bar{F}}_{3}  &  =-\beta\sum_{m}\sum_{i\neq m}\sum_{j\neq m}%
\dfrac{3\left\langle \mathbf{b}_{im}\right\vert \left\vert \mathbf{b}%
_{mj}\right\rangle ^{2}-1}{\left\vert \mathbf{R}_{i}-\mathbf{R}_{m}\right\vert
^{3}\left\vert \mathbf{R}_{j}-\mathbf{R}_{m}\right\vert ^{3}}\label{T6}\\
&  =-\beta\sum_{m}\sum_{i\neq m}\sum_{j\neq m}\dfrac{3\cos^{2}\left(
\measuredangle_{i,m,j}\right)  -1}{\left\vert \mathbf{R}_{i}-\mathbf{R}%
_{m}\right\vert ^{3}\left\vert \mathbf{R}_{j}-\mathbf{R}_{m}\right\vert ^{3}%
}\nonumber
\end{align}
with a prefactor $\beta=3\mu_{0}\chi_{b}^{3}V_{b}^{3}H_{0}^{2}/\left(
64\pi^{2}\right)  $ which is nothing else but Eq.2 from the main text.

Note that for $i\neq j\neq m$ Eq.(S6) contains the three-body interaction
written in Eq.(2) of the manuscript. For $i=j$ it describes the second order
of the two-body interactions which are isotropic (Van der Waals-like) with
$\mathcal{\bar{F}}_{3}^{(i-j)}\sim1/\left\vert \mathbf{R}_{i}-\mathbf{R}%
_{j}\right\vert ^{6}$ (see also Eq.(S5)).

We stress that the above derivations hold for \textit{isotropic} magnetic
beads only, where $(\hat{\chi}_{b})_{\alpha\beta}=\chi_{b}\delta_{\alpha\beta
}$. In the case of anisotropic magnetic objects (e.g. membranes as well as
anisotropic beads) the leading term in $\mathcal{\bar{F}}_{scDI}$ is the
first-order anisotropic dipole-dipole interaction ($1/\left\vert
\mathbf{R}_{i}-\mathbf{R}_{j}\right\vert ^{3}$) and described by Eq.(4) in the
manuscript. Its derivation is given below.

\section{Derivation of the 2-Membrane Interaction (Eq.4)}

In the previous sections we were concerned with interactions of isotropic
spherical particles whose susceptibility tensors were merely diagonal i.e.
$(\hat{\chi}_{b})_{\alpha\beta}=\chi_{b}\delta_{\alpha\beta}$. In the
interesting case of two interacting ellipsoids (spheroids) with orientation
dependent and non-trivial susceptibility tensors $\hat{\chi}_{1},\hat{\chi
}_{2}$, the free-energy in Eq.(1) of the manuscript can be rewritten in the
form
\begin{equation}
\mathcal{\bar{F}}_{scDI}\left(  \mathbf{H}_{0},\{\mathbf{R}_{i}\}\right)
=-\frac{\mu_{0}}{2}H_{0}^{2}V(Tr(\hat{\chi}_{1,eff}+\hat{\chi}_{2,eff}),
\label{M1}%
\end{equation}
where
\begin{equation}
\hat{\chi}_{1,eff}=(1-\varphi_{12}^{2}\hat{\chi}_{1}\hat{t}_{12}\hat{\chi}%
_{2}\hat{t}_{12})^{-1}(\hat{\chi}_{1}-\varphi_{12}\hat{\chi}_{1}\hat{t}%
_{12}\hat{\chi}_{2}) \label{M2}%
\end{equation}
and same for $\hat{\chi}_{2,eff}$ by replacing $1\rightarrow2$. The slightly
more intricate form of $\hat{\chi}_{1/2,eff}$ \ comes now from the fact that
the operator $\hat{t}_{12}=\hat{1}-3\left\vert \mathbf{b}_{12}\right\rangle
\left\langle \mathbf{b}_{12}\right\vert $ and the susceptibilities $\hat{\chi
}_{i}$ don't commute any more. Expanding $\hat{\chi}_{1/2,eff}$ \ to the first
order w.r.t. $\varphi_{12},$ the interaction part of $\mathcal{\bar{F}}%
_{scFI}\left(  \mathbf{H}_{0},\{\mathbf{R}_{i}\}\right)  \equiv\mathcal{\bar
{F}}_{int}\left(  \mathbf{H}_{0},\{\mathbf{R}_{12}\}\right)  $ from main text
simplifies to%

\begin{equation}
\mathcal{\bar{F}}_{int}(1,2)=\frac{\mu_{0}}{2}H_{0}^{2}V\varphi_{12}%
Tr\{(\hat{\chi}_{1}\hat{\chi}_{2}+\hat{\chi}_{2}\hat{\chi}_{1})\hat{t}_{12}\}.
\label{M3}%
\end{equation}
Having in mind two identical membranes (with volume $V_{m}$) we consider
susceptibilities $\hat{\chi}_{1}$ and $\hat{\chi}_{2}$ of two oblate
spheroids, which are differently oriented in space. In terms of their own
local coordinate systems (in Dirac bra-ket notation for tensors) they are
given by%
\begin{align}
\hat{\chi}_{1}  &  =\chi_{\min}\left\vert \mathbf{n}_{1}\right\rangle
\left\langle \mathbf{n}_{1}\right\vert +\chi_{\max}(\hat{1}-\left\vert
\mathbf{n}_{1}\right\rangle \left\langle \mathbf{n}_{1}\right\vert
)\label{M4}\\
\hat{\chi}_{2}  &  =\chi_{\min}\left\vert \mathbf{n}_{2}\right\rangle
\left\langle \mathbf{n}_{2}\right\vert +\chi_{\max}(\hat{1}-\left\vert
\mathbf{n}_{2}\right\rangle \left\langle \mathbf{n}_{2}\right\vert ),\nonumber
\end{align}
where the unit vectors $\left\vert \mathbf{n}_{1}\right\rangle ,$ $\left\vert
\mathbf{n}_{2}\right\rangle $ are the normals of the membranes $1$ and $2$
respectively. By using Eq.(\ref{M4}), and noting that $Tr\{\hat{\chi}_{1}%
\hat{\chi}_{2}\}=$ $Tr\{\hat{\chi}_{2}\hat{\chi}_{1}\}$, $Tr\{\hat{\chi}%
_{1}\hat{\chi}_{2}(\left\vert \mathbf{b}_{21}\right\rangle \left\langle
\mathbf{b}_{21}\right\vert )\}=Tr\{\hat{\chi}_{2}\hat{\chi}_{1}(\left\vert
\mathbf{b}_{12}\right\rangle \left\langle \mathbf{b}_{12}\right\vert )\}$ and
$Tr\{\left\vert \mathbf{n}_{i}\right\rangle \left\langle \mathbf{n}%
_{j}\right\vert \}=\mathbf{\langle n}_{i}\left\vert \mathbf{n}_{j}%
\right\rangle $ (with $\mathbf{a}\cdot\mathbf{b\equiv\langle a}\left\vert
\mathbf{b}\right\rangle $ the scalar product) it follows
\begin{equation}
Tr\{\hat{\chi}_{1}\hat{\chi}_{2}\}=\chi_{\max}^{2}\left[  1+2\gamma+c_{3}%
^{2}(1-\gamma)^{2}\right]  \label{M5}%
\end{equation}%
\begin{align}
Tr\{\hat{\chi}_{1}\hat{\chi}_{2}(\left\vert \mathbf{b}_{12}\right\rangle
\left\langle \mathbf{b}_{12}\right\vert )\}  &  =\chi_{\max}^{2}%
[1-(1-\gamma)(c_{1}^{2}+c_{2}^{2})\label{M6}\\
&  +(1-\gamma)^{2}c_{1}c_{2}c_{3}],\nonumber
\end{align}
where $\gamma=(\chi_{\min}/\chi_{\max})$ and $c_{1}=\mathbf{n}_{1}%
\cdot\mathbf{b}_{12}$, $c_{2}=\mathbf{n}_{2}\cdot\mathbf{b}_{12}$,
$c_{3}=\mathbf{n}_{1}\cdot\mathbf{n}_{2}$ are factors describing the mutual
orientation of membranes. By replacing Eqs.(\ref{M5}-\ref{M6}) in
Eq.(\ref{M3}) (where $V_{b}$ in $\varphi_{12}$ is replaced by the membrane
volume $V_{m}$) one obtains Eq.(4) in the manuscript.

\section{Derivation of the equation of state $p(f_{V})$ \ (Eq.5)}

Let us define the volume fraction $f_{V}=(N_{m}V_{m}/V)\approx V_{b}^{tot}/V$
where $V_{m}(\approx Da_{M}^{2})$ is the volume of the single membrane and
$N_{m}$ is the total number of (equal) membranes in the container volume
$V\approx Na_{M}^{3}$, and $V_{b}^{tot}$ is the total volume of the beads.
Here, $D$ is the bead diameter and $a_{M}$ is the size of the single membrane
- see Fig.2c in the manuscript. It follows that $f_{V}\approx3D/a_{M}$. Note,
that in the following we fix $V_{m}^{tot}=N_{m}V_{m}$, i.e. $V_{m}%
^{tot}=const$.

The pressure is defined by $p=-\partial\mathcal{\bar{F}}^{tot}/\partial V$
where $\mathcal{\bar{F}}^{tot}\mathcal{=\bar{F}}_{self}^{tot}\mathcal{+\bar
{F}}_{int}^{tot}$ is the total energy of the membranes, $\mathcal{\bar{F}%
}_{self}^{tot}$ is the self-energy of (non-interacting) membranes and
$\mathcal{\bar{F}}_{int}^{tot}$ is the interaction energy of membranes. $Eq.3$
of the manuscript gives the energy of the single membrane while the total
self-energy of $N_{m}$ membranes ($V_{m}^{tot}=N_{m}V_{m}$) is given by
\begin{equation}
\mathcal{\bar{F}}_{self}^{tot}=-(2\chi_{\max}+\chi_{\min})V_{m}^{tot}%
\frac{B_{0}^{2}}{2\mu_{0}}\label{p1}%
\end{equation}%
\begin{equation}
\chi_{\max}=\frac{\chi}{1+L_{m}\chi},\text{ }\chi_{\min}=\frac{\chi
}{1+(1-2L_{m})\chi}\label{p2}%
\end{equation}
for simplicity, we study only the case with $\chi\gg1$ (note the material
susceptibility fulfills $\chi>\chi_{b}\leq3$, $\chi_{b}$ is the bead
susceptibility with respect to the applied (external) field) and $L_{m}\chi
\ll1$ ($L_{m}\ll1$). Since a membrane is considered as an extreme oblate
ellipsoid with semi-axes $c\approx D/2$, $a=b\approx a_{M}/2$ one obtains for
$L_{m}\approx f_{V}/4$ (i.e. $f_{V}\ll1$) - for the expression for $L_{m}$ in
terms of $a,c$ see Ref.13 of the manuscript. After a straightforward small
$L_{m}$ expansion one obtains%
\begin{equation}
\mathcal{\bar{F}}_{self}^{tot}(V)\approx-(const+\frac{1}{2}f_{V}\chi^{2}%
)V_{m}^{tot}\frac{B_{0}^{2}}{2\mu_{0}},\label{p3}%
\end{equation}
where $const$ is independent of $f_{V}$ (note, $f_{V}=(V_{m}^{tot}/V)$ and
$V_{m}^{tot}=const$) .

The total interaction energy of membranes $\mathcal{\bar{F}}_{int}^{tot}$ is
\begin{equation}
\mathcal{\bar{F}}_{int}^{tot}(V)=\frac{1}{2}\sum_{i,j}\mathcal{\bar{F}}%
_{int}(i,j), \label{p4}%
\end{equation}
where the pair-interaction energy $\mathcal{\bar{F}}_{int}(i,j)$ is given by
$Eq.4$ of the manuscript - its derivation is given above. For $\chi\gg1$ one
obtains
\begin{equation}
\mathcal{\bar{F}}_{int}^{tot}(V)\approx(\frac{1}{8\pi}f_{V}\chi^{2}%
S)V_{m}^{tot}\frac{B_{0}^{2}}{2\mu_{0}}, \label{p5}%
\end{equation}
where the constant $S$ is given by a sum over all membranes $i$ in the lattice
interacting with a given membrane (with index $j$)
\begin{equation}
S=\sum_{i=1,i\neq j}^{N}\left(  C_{1}^{2}\left(  i,j\right)  +C_{2}^{2}\left(
i,j\right)  +\frac{1}{3}C_{3}^{2}\left(  i,j\right)  -C_{1}\left(  i,j\right)
C_{2}\left(  i,j\right)  C_{3}\left(  i,j\right)  -\frac{2}{3}\right)
\frac{a_{M}^{3}}{\left\vert \mathbf{R}_{ji}\right\vert ^{3}} \label{p6}%
\end{equation}
with geometric factors $C_{1}\left(  i,j\right)  =\mathbf{n}_{i}%
\cdot\mathbf{b}_{ij}$, $C_{2}\left(  i,j\right)  =\mathbf{n}_{j}%
\cdot\mathbf{b}_{ij}$, $C_{3}\left(  i,j\right)  =\mathbf{n}_{i}%
\cdot\mathbf{n}_{j}$ (also given below Eq.4 in the manuscript). For the
definition of $\mathbf{n}_{1,2}$ and $\mathbf{b}_{12}$ see above the
derivation of Eq.4. \ If we choose the membrane $j$ (without restriction)
arbitrarily in the $z$-direction, i.e. $\mathbf{n}_{j}=\left(  0,0,1\right)
$, the lattice sum $S$ can be practically decomposed into 3 terms%

\[
S=S_{x}+S_{y}+S_{z}%
\]
each representing the contribution of one of the 3 sublattices (in $x$ , $y$
and $z$ direction). By evaluating the scalar products appearing in
$C_{1/2/3}\left(  i,j\right)  $ at each lattice site membrane $i,$ the
sub-lattice sums can be written as:%

\begin{align*}
S_{z}  &  =\sum_{\left(  l_{x},l_{y},l_{z}\right)  \neq\left(  0,0,0\right)
}\frac{1}{\left(  l_{x}^{2}+l_{y}^{2}+l_{z}^{2}\right)  ^{3/2}}\left(
\frac{l_{z}^{2}}{l_{x}^{2}+l_{y}^{2}+l_{z}^{2}}-\frac{1}{3}\right) \\
S_{x}  &  =S_{y}=\sum_{\left(  l_{x},l_{y},l_{z}\right)  \neq\left(
0,0,0\right)  }\frac{1}{\left(  l_{x}^{2}+l_{y}^{2}+l_{z}^{2}\right)  ^{3/2}%
}\left(  \frac{l^{2}+j^{2}}{l_{x}^{2}+l_{y}^{2}+l_{z}^{2}}-\frac{2}{3}\right)
\end{align*}
where the sums run over all integer triplets $\left(  l_{x},l_{y}%
,l_{z}\right)  $ (from $-\infty$ to $+\infty$) which don't vanish all at the
same time. A numerical summation of these 3 terms $S_{x/y/z}$ gives finally a
value $S\approx10.$

Summing up the results above, the pressure is finally given by
\begin{align}
p  &  =-\frac{\partial\mathcal{\bar{F}}^{tot}}{\partial V}=(\frac{1}{2}%
+\frac{S}{8\pi})\chi^{2}f_{V}^{2}\frac{B_{0}^{2}}{2\mu_{0}}\label{p7}\\
&  \approx0.5\chi^{2}f_{V}^{2}\frac{B_{0}^{2}}{\mu_{0}}%
\end{align}